\documentclass[12pt]{article}
\usepackage{a4,t1enc,amsmath,curves}
\usepackage[dvips]{graphicx}
\makeatother
\usepackage[ps,line,frame,import,arrow]{xy}
\xyoption{dvips}
\makeatletter
\DeclareGraphicsExtensions{ps,eps}
\def\newpic#1{\def\emline##1##2##3##4##5##6{\put(##1,##2){\special{em:point #1##3}}
\put(##4,##5){\special{em:point #1##6}}
\special{em:line #1##3,#1##6}}}
\newpic{}

\begin{document}
%\baselineskip=2.\normalbaselineskip

\title{Can spontaneous symmetry breaking occur in potential with one minima?}
\author{Art\=uras Acus}
\date{}
\maketitle

\begin{center}
\it Institute of Theoretical Physics and Astronomy,
Vilnius, 2600 Lithuania
\\ {and}\\ Vilnius Pedagogical University, Vilnius, 2600 Lithuania
\end{center}

{\it Introduction.} Spontaneous symmetry breaking occurs when the
symmetry that a physical system possesses, is not preserved for the ground
state of the system. Although the procedure of symmetry breaking
is quite clear from the mathematical point of view, the physical
interpretation of the phenomenon is worth to be better understood. In this
note we present a simple and instructive example of the symmetry breaking in
a mechanical system. It demonstrates that the spontaneous symmetry breaking
can occur for the spatially extended solutions in a potential characterised
by a single minimum.

A history of the example is as follows. The pupils were suggested the
following exercise at the National Lithuanian olympiad in physics a few
years ago. The system considered is a massive square situated in a sharp
angle of  $\Theta =60^{\circ }$ (Figure~1), the square being affected by the
homogeneous Earth field. One is required to find the configuration, for
which the potential energy of the massive square is minimum, i.e to find the
ground state of the system. The answer to the exercise was provided by the
exercise book${}^1$. To great surprise of the
examiners a number of pupils came to a quite different answer which proved
to be the only correct solution. So the answer provided by the exercise
book appeared to be wrong.  

{\it Experiments.} Before starting mathematical investigation, let us
make a little experiment. Let us take a (hardcover) book, and any cube, such
as Rubbick cube or a dice. Let us slightly open the cover and drop the cube
into the emerging gap. What kind of position the cube
took?

There is another way to solve the exercise ''experimentally'' that has been
found by a quick-witted pupil. Let us cut a square from a sheet of paper and
make a pinhole in the center of the square. Put a spike of a pencil in the
hole and turn slowly the square keeping it inside the angle drawn on another
sheet of paper. This results in a nice closed curve which can be used to
determine the minimum position of the center of mass.

{\it The solution.} Let us now formulate the problem
mathematically, not restricting ourself to a particular value of the angle $%
\Theta $. When the angle is sharp, one can see readily two possible
positions of the square, where either the diagonal (Figure~2a) or the edge
(Figure~2b) of the square touches the sides of the angle $\Theta $. (The
third trivial possibility arises when angle $\Theta $ is obtuse.)

\newsavebox{\toto }
\savebox{\toto }[6.5cm]{
\unitlength=1.1mm
\special{em:linewidth 0.01pt}
\linethickness{0.01pt}
\begin{picture}(30.00,40.00)
\put(0.,0.){\circle*{0.3}}
\emline{00.00}{00.00}{51}{20.00}{30.00}{52}
\emline{00.00}{00.00}{53}{-20.00}{30.00}{54}
\emline{00.00}{00.00}{55}{00.00}{34.00}{56}
\emline{-20.00}{00.00}{57}{20.00}{00.00}{58}
\emline{-8.00}{12.00}{59}{10.00}{15.00}{60}
\emline{-8.00}{12.00}{61}{-11.00}{30.00}{62}
\put(10.0,15){\circle*{0.3}}
\put(-8.0,12){\circle*{0.3}}
\put(10.0,15){\arc(-3.,-4.5){-47.}}
\put(-8.0,12){\arc(2.2,-3.3){66.}}
\emline{10.00}{15.00}{63}{7.00}{33.00}{64}
\emline{-11.00}{30.00}{65}{7.00}{33.00}{66}
\emline{-8.00}{12.00}{67}{7.00}{33.00}{68}
\emline{-8.00}{12.00}{89}{-8.50}{12.00}{90}
\emline{-8.50}{22.50}{89}{2.00}{22.50}{90}
\emline{-11.00}{30.00}{69}{10.00}{15.00}{70}
\emline{-0.50}{22.50}{71}{-0.50}{.90}{72}
\put(-0.50,0.90){\circle*{0.3}}
\emline{-8.00}{12.00}{73}{-8.00}{00.00}{74}
\emline{-8.00}{12.00}{73}{-8.00}{24.00}{74}
\emline{-8.00}{12.00}{75}{-8.50}{10.50}{76}
\emline{-8.00}{12.00}{77}{-8.50}{13.50}{78}
\emline{-8.00}{0.00}{77}{-8.50}{1.50}{78}
\emline{-8.00}{0.00}{79}{-7.50}{1.50}{80}
\emline{-8.00}{12.00}{81}{-7.50}{13.50}{82}
\emline{-8.00}{12.00}{83}{-7.50}{10.50}{84}
\emline{-8.00}{22.50}{85}{-7.50}{21.00}{86}
\emline{-8.00}{22.50}{87}{-8.50}{21.00}{88}
%\emline{-0.50}{10.50}{75}{-0.50}{9.00}{76}
%\emline{-0.50}{7.50}{85}{-0.50}{6.00}{86}
%\emline{-0.50}{4.50}{79}{-0.50}{3.10}{80}
%\emline{-0.50}{17.00}{77}{2.00}{18.00}{78}
%\emline{-0.50}{1.50}{83}{-2.00}{-1.00}{84}
\put(0.60,3.00){\makebox(0,0)[cc]{$\Theta $}}
\put(0,0){\arc(3.,4.5){67.38}}
\put(0.50,-2.00){\makebox(0,0)[cc]{C}}
\put(0.00,-8.00){\makebox(0,0)[cc]{}}
\put(-1.70,18.50){\makebox(0,0)[cc]{$\alpha $}}
\put(-0.5,22.5){\circle*{0.3}}
\put(-0.5,22.5){\arc(-0.3,-2.5){-29.8}}
\put(-0.70,25.50){\makebox(0,0)[cc]{E}}
\put(-3.,1.00){\makebox(0,0)[cc]{D}}
\put(12.50,15.00){\makebox(0,0)[cc]{B}}
\put(10.00,23.50){\makebox(0,0)[cc]{$a$}}
\put(7.50,13.50){\makebox(0,0)[cc]{$\Phi $}}
\put(-4.00,7.50){\makebox(0,0)[cc]{$\ell $}}
\put(-6.00,10.80){\makebox(0,0)[cc]{$\Psi $}}
\put(-10.00,12.00){\makebox(0,0)[cc]{A}}
\put(-10.00,5.00){\makebox(0,0)[cc]{$h^{\prime }$}}
\put(-10.00,17.50){\makebox(0,0)[cc]{$h^{\prime \prime }$}}
\end{picture}
} 
\newsavebox{\tutu }
\savebox{\tutu}[6.5cm]{
\unitlength=1.1mm
\special{em:linewidth 0.01pt}
\linethickness{0.01pt}
\begin{picture}(30.00,40.00)
\put(0.,0.){\circle*{0.3}}
\put(0,0){\arc(3.,4.5){67.38}}
\emline{00.00}{00.00}{1}{20.00}{30.00}{2}
\emline{00.00}{00.00}{3}{-20.00}{30.00}{4}
\emline{00.00}{00.00}{5}{00.00}{34.00}{6}
\emline{-20.00}{00.00}{7}{20.00}{00.00}{8}
\emline{-13.81}{20.72}{9}{0.50}{32.04}{10}
\emline{0.50}{32.04}{11}{11.82}{17.72}{12}
\emline{11.82}{17.72}{13}{-2.50}{6.41}{14}
\emline{-2.50}{6.41}{15}{-13.81}{20.72}{16}
\emline{-13.81}{20.72}{17}{11.82}{17.72}{18}
\put(11.82,17.72){\circle*{0.3}}
\put(11.82,17.72){\arc(-3.,-4.5){-63.38}}
\put(-13.81,20.72){\circle*{0.3}}
\put(-13.81,20.72){\arc(2.2,-3.3){50.}}
\emline{0.50}{32.04}{19}{-2.50}{6.41}{20}
\emline{-1.00}{19.22}{21}{-1.00}{00.00}{22}
% \emline{-1.00}{1.50}{29}{-4.00}{1.50}{30}
\emline{17.00}{19.22}{23}{17.00}{00.00}{24}
\emline{17.00}{19.22}{43}{16.50}{17.78}{44}
\emline{17.00}{19.22}{45}{17.50}{17.78}{46}
\emline{17.00}{0.00}{47}{16.50}{1.5}{48}
\emline{17.00}{0.00}{49}{17.50}{1.5}{50}
\emline{-4.00}{19.22}{25}{18.50}{19.22}{26}
\emline{-1.00}{15.22}{27}{-3.00}{18.22}{28}
\emline{-13.81}{20.72}{33}{-13.81}{00.00}{34}
\emline{11.82}{17.72}{31}{11.82}{00.00}{32}
\put(-1.0,1.60){\circle*{0.3}}
\put(-1.0,19.22){\circle*{0.3}}
\put(-1.0,19.22){\arc(-0.1,-5.3){-5}}
\put(0.50,3.00){\makebox(0,0)[cc]{$\Theta $}}
\put(-3.50,18.00){\makebox(0,0)[cc]{$\alpha $}}
\put(9.00,15.50){\makebox(0,0)[cc]{$\Phi $}}
\put(14.00,17.50){\makebox(0,0)[cc]{D}}
\put(11.82,-3.00){\makebox(0,0)[cc]{G}}
\put(-13.81,-3.00){\makebox(0,0)[cc]{F}}
\put(0.00,-3.00){\makebox(0,0)[cc]{C}}
\put(7.50,25.50){\makebox(0,0)[cc]{$a$}}
\put(-0.70,22.00){\makebox(0,0)[cc]{E}}
\put(-15.50,19.50){\makebox(0,0)[cc]{A}}
\put(-9.00,17.50){\makebox(0,0)[cc]{$\Psi $}}
\put(-4.00,2.00){\makebox(0,0)[cc]{B}}
\put(15.00,7.00){\makebox(0,0)[cc]{$h_a$}}
\put(0.00,-8.00){\makebox(0,0)[cc]{}}
\put(7.00,8.00){\makebox(0,0)[cc]{$f$}}
\put(-7.00,8.00){\makebox(0,0)[cc]{$\ell $}}
\put(-6.00,22.00){\makebox(0,0)[cc]{$d$}}
\end{picture}
} 

%\begin{figure}
%\label{possiblepositions}
%\caption{possible}
%\begin{center}
%\usebox{\tutu } \usebox{\toto }\vspace{1cm}
%\end{center}
%\end{figure}
%\vspace{1cm}

To determine the orientation of the object inside the angle $\Theta $, let
us define the angle $\alpha $ between the diagonal and a vertical line going
from the center of the square (Figures 2a, 2b). The vertical do not
generally coincide with the bisector of the angle $\Theta $. Potential
energy of the homogeneous massive square can be written as: $V=mgh$, where $h
$ is the height of the center of the square. The problem thus reduces to
finding the quantity $h$ as a function of two angles ($\Theta $ and $\alpha $%
) and the length of the square edge $a$. Three cases mentioned above should
be treated separately.

\begin{description}
\item {\bf a.} The diagonal touches the angle sides (Figure~2a), i.e. $%
\frac{\pi }{4}+\frac{\Theta }{2}\leq \alpha \leq \frac{\pi }{4}-\frac{\Theta 
}{2}$ and $0<\Theta \leq \frac{\pi }{2}.$ Analyzing $\triangle ABE$ we have: 
$\Psi =\frac{\pi }{2}-\alpha -\frac{\Theta }{2}$. Similarly $\triangle ACD$
yields: $\Psi =\pi -\Phi -\Theta $. Thus one finds:\ $\Phi =\frac{\pi }{2}%
+\alpha -\frac{\Theta }{2}$. Applying the sine theorem to triangle $ACD$ we
get: $\frac{d}{\sin \Theta }=\frac{f}{\sin \Psi }$ and $\frac{d}{\sin \Theta 
}=\frac{\ell }{\sin \Phi }$. Therefore we can express $f$ and $\ell $
through the angles $\alpha ,\Theta $ and the parameter $d=a\sqrt{2}$.
Finally, the height $h$ can be found as an average of altitudes of the
trapezoid $AFGD$: 
\[
h_{a}=\frac{1}{2}\left( f\sin \frac{\pi -\Theta }{2}+\ell \sin \frac{\pi
-\Theta }{2}\right) =\frac{1}{\sqrt{2}}\cos \alpha \cot \frac{\Theta }{2}.
\]

\item  {\bf b.} The edge touches the angle sides (Figure~2b): $0<\Theta
\leq \frac{\pi }{2}$ and $\frac{\pi }{4}-\frac{\Theta }{2}\leq \alpha \leq 
\frac{\pi }{4}$. In this case the total height $h_{b}$ can be calculated as
a sum: $h_{b}=$ $h^{\prime }+h^{\prime \prime }$. Analyzing $\triangle ABC$
we have $\Theta +\Phi +\Psi =\pi $. The triangle $\triangle ADE$ yields: $%
\frac{\pi }{4}+\Psi +\frac{\Theta }{2}+\alpha =\pi $, so that $\Phi =\frac{%
\pi }{4}+\alpha -\frac{\Theta }{2}$. Applying the sine theorem to the
triangle $ABC$ yields: $\frac{a}{\sin \Theta }=\frac{\ell }{\sin \Phi }$.
Now we can express $\ell $, giving: $h^{\prime }=\ell \cos \frac{\Theta }{2}.
$ On the other hand, one finds readily: $h^{\prime \prime }=\frac{\sqrt{2}}{2%
}a\cos \alpha $. As a result,\ the total height takes the form:
\[
h_{b}=a\frac{\cos \frac{\Theta }{2}}{\sin \Theta }\sin (\frac{\pi }{4}%
+\alpha -\frac{\Theta }{2})+\frac{\sqrt{2}}{2}a\cos \alpha .
\]
The same procedure can be carried out for $-\frac{\pi }{4}\leq \alpha \leq -%
\frac{\pi }{4}+\frac{\Theta }{2}$.

\item  {\bf c.} The angle $\Theta $ is obtuse: $\frac{\pi }{2}\leq \Theta
\leq \pi $ and $-\frac{\pi }{4}+\frac{\Theta }{2}\leq \alpha \leq \frac{\pi 
}{4}-\frac{\Theta }{2}$. In this case the height $h_{c}$ depends only on the
angle $\alpha $: 
\[
h_{c}=a\frac{\sqrt{2}}{2}\cos \alpha .
\]
\end{description}

Results are plotted for two different sharp angles: $\Theta =\pi /6$
(Figures~3a) and $\Theta =\pi /3$ (Figures~3b). The most fascinating feature
is that the energy  minimum of the system appears in the positions (2) and
(4) which break the initial symmetry of the system, rather than in the
positions positions (1), (5) or (3) that preserve this symmetry. In the
cases (2) and (4), symmetry becomes broken spontaneously, because we cannot
determine which particular position of the square (left (2) or right
(4)) will appear. The phenomenon does not depend on a particular
value of the\ angle $\Theta $, as one can see from Figure~4 showing clearly
two ''vacua'' valleys. The only exception is the point $\Theta =90^{\circ }
$, when both configurations coincide. Figures 3-4 have been plotted using
''Mathematica''${}^2$ (Wolfram Research).

What are the differences between well known example of 'mexican hat' (i.e.
the central cut of hat if we restrict ourselves to two dimensions) and 
our example? In the case of the 'mexican hat', one has a point like particle
in potential with two equivalent minima. In contrast, here we are dealing
with a spatially extended object (the square) residing in a potential
characterized by a single minimum (the angle). If we regard the square as
an energy density of a soliton-like solution (similar to those fascinating
shapes${}^{3,4}$ recently discovered in the Skyrme
model), then our example suggests that the spontaneous symmetry breaking can
occur in a potential with a single minimum for extended (soliton-like)
solutions.

{\it Conclusions.}
We hope that the example considered will help teachers
to explain the spontaneous symmetry breaking and inspire pupils to look for
manifestation of his phenomenon in three dimensional systems. The example
can be easily analyzed and demonstrated in simple experiments. Yet it
provides far reaching conclusions.

{\it Acknowledgment.} I would like to thank dr. V. \v{S}imonis for
pointing me out this exercise and especially professors P. Bogdanovi\v{c}ius
and S. Ali\v sauskas
both for helping me to locate the exercise book${}^1$
and drawing my attention to the second ''experimental'' solution.

{\bf References}\newline
${}^1$ Iosif Sh. Slobodecki and Vladimir A. Orlov, {\it
Vsesoyuznye olimpiady po fizike} (Moskva, Prosveshchenie,1982), pp. 67.\newline
${}^2$ Stephen Wolfram,{\it The Mathematica Book} (Wolfram Media, Cambridge University Press, 1999), 4th ed., pp. 473--557.\newline
${}^3$ Wojtek Zakrzewski, "Models link nuclei with buckyballs", Physics World {\bf 10} (11), 26-27 (1997).\newline
${}^4$ C.J. Houghton, N.S. Manton, and P.M. Sutcliffe "Rational maps, monopoles and skyrmions", Nucl. Phys. B {\bf 510}, 507--537 (1998).
\newpage
\begin{center}
List of figures
\end{center}
\begin{description}
\item Figure~1. The axially symmetric system.
\item Figure~2a. Square position: diagonal touches angle sides.
\item Figure~2b. Square position: edge touches angle sides. 
\item Figure~3a. Potential energy of the square for $\Theta=\frac{\pi}{6}$.
\item Figure~3b. Potential energy of the square  for $\Theta=\frac{\pi}{3}$.
\item Figure~4. "Vacua" valleys of the system.
\end{description}
\newpage
\thispagestyle{empty}
Figure~1. 

\vspace{7cm}

$$
\xy
(0,0)="0" ;
"0"+/r2.4pc/+(0,20);
**@{-};
"0"-/r2.4pc/+(0,20);
**@{-};
"0" *-/3pc/!D{\scriptstyle 60^\circ};
"0"+(0,20) *=<1cm,1cm>\frm{*};
(0,9), \ar@*{[|(2)]}(0,14);
\endxy
$$

\newpage
\thispagestyle{empty}
Figure~2a.
\vspace{7cm}

\begin{center}
\usebox{\tutu }
\end{center}

\newpage
\thispagestyle{empty}
Figure~2b.
\vspace{7cm}

\begin{center}
\usebox{\toto }
\end{center}

\newpage
\thispagestyle{empty}
{Figure~3a.}
\vspace{7cm}

$$
\xy
\xyimport(3.7,3.7)(1.4,1.4){
\includegraphics*[width=7.5cm]{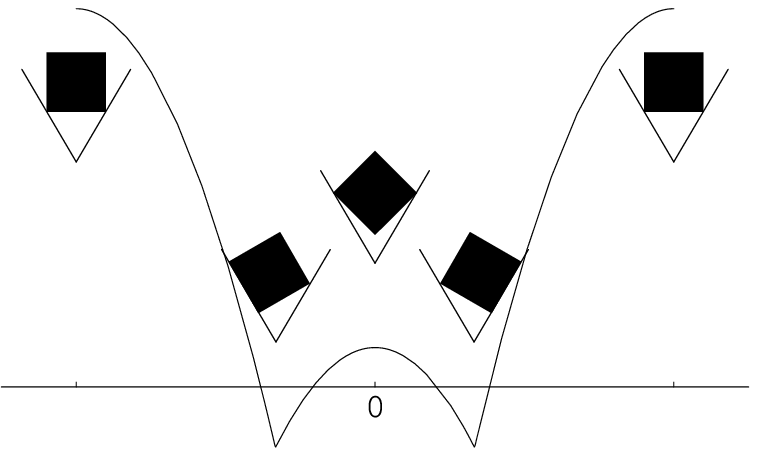}}
*+!U{},
(0.75,-1.7)*+!RD{\Theta=\frac{\pi}{6}},
(2.2,-1.2)*+!LD{\alpha},
(1.84,-1.4)*+!LD{\frac{\pi}4},
(-1.2,-1.4)*+!LD{-\frac{\pi}4},
(0.9,0.5)*+!LD{4},
(0.4,1.4)*+!LD{3},
(-1.04,0.3)*+!LD{1},
(-0.1,0.5)*+!LD{2},
(1.85,0.3)*+!LD{5},
\endxy
$$

\newpage
\thispagestyle{empty}
{Figure~3b.}
\vspace{7cm}

$$
\xy
\xyimport(3.7,3.7)(1.4,1.4){
\includegraphics*[width=7.5cm]{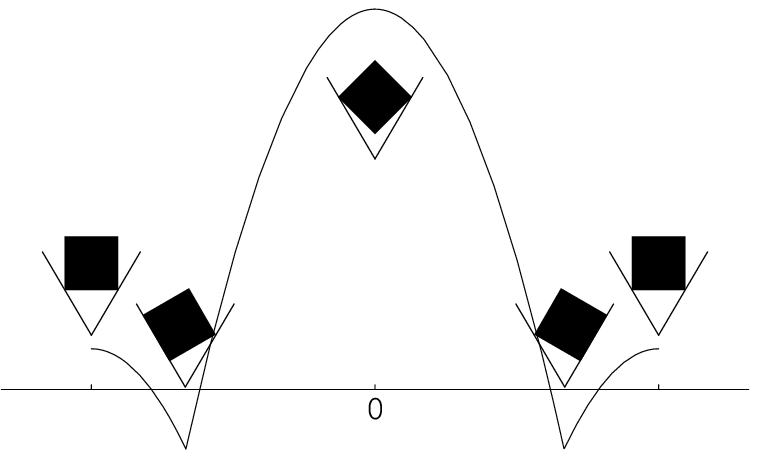}}
*+!U{},
(0.75,-1.7)*+!RD{\Theta=\frac{\pi}{3}},
(2.2,-1.2)*+!LD{\alpha},
(1.75,-1.4)*+!LD{\frac{\pi}4},
(-1.15,-1.4)*+!LD{-\frac{\pi}4},
(1.3,0.2)*+!LD{4},
(0.39,.4)*+!LD{3},
(-1.,0.6)*+!LD{1},
(-0.5,0.2)*+!LD{2},
(1.75,0.6)*+!LD{5},
\endxy
$$

\newpage
\thispagestyle{empty}
{Figure~4.}
\vspace{7cm}

\begin{center}
\xy
\xyimport(3.7,3.7)(1.4,1.4){
\includegraphics*[width=8.cm,height=7.cm]{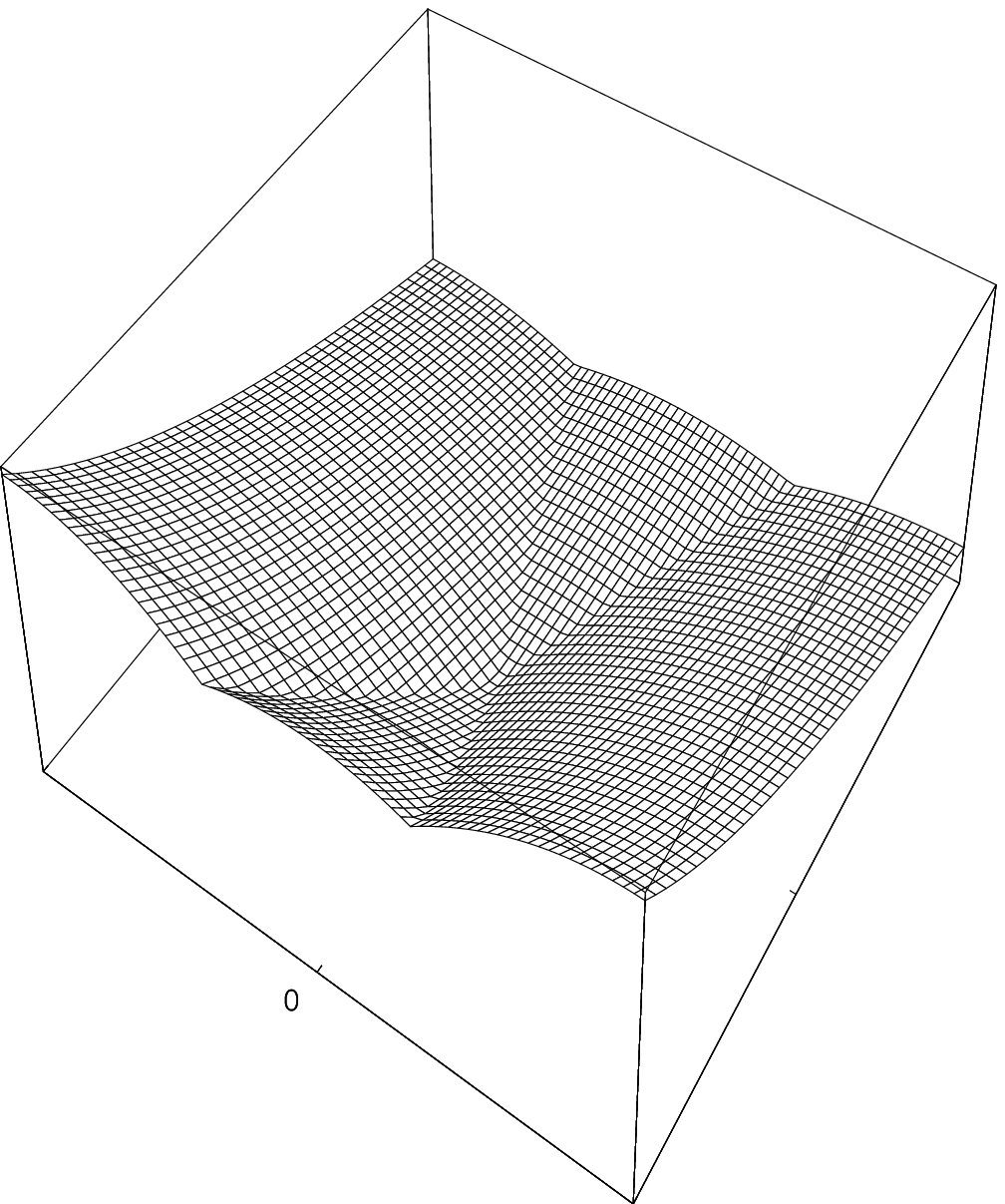}}
*+!U{}, (-1.3,0.4)*+!R{h},
(-1.1,-0.3)*+!RD{-\frac{\pi}{4}},
(0.81,-1.56)*+!RD{\frac{\pi}{4}},
(1.1,-1.5)*+!LD{\frac{\pi}{3}},
(-0.33,-1.05)*+!LD{\alpha},
(1.63,-0.66)*+!LD{\frac{\pi}2},
(2.1,0.2)*+!LD{\frac{\pi}{\sqrt{2}}},
(1.7,-0.86)*+!LD{\Theta},
\endxy
\end{center}

\end{document}